\documentclass[]{aastex631}

\usepackage{amsmath} 
\usepackage{amsfonts,amssymb} 
\usepackage{gensymb} 
\usepackage{natbib}

\begin{document}
	
	\title{Spectro-Polarimetric Properties of Sunquake Sources in X1.5 Flare and Evidence for Electron and Proton Beam Impacts}
	
\shorttitle{Sunquake Initiation} 
\shortauthors{Kosovichev et al.}
	
	\author[0000-0003-0364-4883]{Alexander G. Kosovichev} \affiliation{Department of Physics, New Jersey Institute of Technology, Newark, NJ 07102} \affiliation{NASA Ames Research Center, Moffett Field, Mountain View, CA 94040}
	
	\correspondingauthor{Alexander G. Kosovichev} \email{alexander.g.kosovichev@njit.edu}
	
	\author[0000-0002-4001-1295]{Viacheslav M. Sadykov} \affiliation{Physics \& Astronomy Department, Georgia State University, Atlanta, GA 30303}

	\author[0000-0002-5519-8291]{John T. Stefan} \affiliation{Department of Physics, New Jersey Institute of Technology, Newark, NJ 07102}

	\begin{abstract} The first significant sunquake event of Solar Cycle 25 was observed during the X1.5 flare of May 10, 2022, by the Helioseismic and Magnetic Imager (HMI) onboard the Solar Dynamics Observatory. We perform a detailed spectro-polarimetric analysis of the sunquake photospheric sources, using the Stokes profiles of the Fe~I 6173~\AA~ line, reconstructed from the HMI linear and circular polarized filtergrams. The results show fast variations of the continuum emission with rapid growth and slower decay lasting 3-4 min, coinciding in time with the hard X-ray impulses observed by the Konus instrument onboard the Wind spacecraft. The variations in the line core appeared slightly ahead of the variations in the line wings, showing that the heating started in the higher atmospheric layers and propagated downward. The most significant feature of the line profile variations is the transient emission in the line core in three of the four sources, indicating intense, impulsive heating in the lower chromosphere and photosphere. In addition, the observed variations of the Stokes profiles reflect transient and permanent changes in the magnetic field strength and geometry in the sunquake sources. Comparison with the radiative hydrodynamics models shows that the physical processes in the impulsive flare phase are substantially more complex than those predicted by proton and electron beam flare models currently presented in the literature. \end{abstract} 
	
	\section{Introduction}

	Sunquakes represent the helioseismic response to solar flares. Acoustic waves, excited by strong photospheric impacts, travel through the solar interior and reappear on the surface as expanding circular-shaped ripples \citep{Kosovichev1998}. The velocity amplitude, observed by the Michelson Doppler Imager (MDI) onboard Solar and Heliospheric Observatory (SoHO), reaches several hundred meters per second, and the wavefront speed increases with the distance, typically from 10 km/s to 100 km/s. The observed characteristics of sunquakes fit very well with the theory of helioseismic waves \citep{Stefan2020}. However, the origin of the strong photospheric perturbations -- sunquake sources -- is not yet understood. Thus, sunquakes represent a major unsolved problem of solar flare physics.
	
	It has been established that sunquakes are usually observed in compact impulsive flares (or compact impulsive parts of complex flares). The sunquake sources are associated with hard X-ray impulses and are observed by the SoHO/MDI and SDO/HMI instrument as transient photospheric perturbations of Doppler velocity, magnetic field, and continuum ("white-light") intensity, occupying areas of just a few pixels \citep[with the total angular size of 1-2 arcsec, e.g.,][]{Kosovichev2006,Sharykin2018,CastellanosDuran2018}. {The impulsive bright and compact white-light flare kernels have been a subject of numerous previous investigations \citep[e.g.,][]{Slonim1975,Babin1999,Matthews2003,Wang2009}.}

	Among the 500 M-X class flares observed by HMI during Solar Cycle 24 (excluding the limb and near-limb flares), 94 helioseismic events were detected \citep{Sharykin2014}. The statistical analysis showed that the sunquake power correlates with the maximum value of the soft X-ray flux time derivative, confirming that the impacts of high-energy particles cause the sunquakes.

	The initial theoretical modeling in the framework of the electron-beam driven `thick-target'  model by \citet{Kosovichev1995} predicted the helioseismic wave amplitude on the solar surface of about several m/s. Non-LTE modeling of the Ni~I 6768~\AA~ line, observed by the MDI instrument, predicted decreases of the line core depth by about 30\% during the electron beam heating and changes in the emission for a few seconds, which was difficult to detect with the 1-min cadence of the MDI instrument \citep{Zharkova2002}.
	
	Further non-LTE modeling of the SDO/HMI observables by \citet{Sadykov2020}, based on the RADYN flare hydrodynamics models \citep{Allred2015} from the F-CHROMA project \citep{Carlsson2023}, found a substantially weaker impact of the electron-beams on the low atmosphere. This study showed that the electron-beam `thick-target' model \citep{Brown1971,Hudson1972}, often considered the standard flare model, has some challenges with explaining the sunquakes and the observed variations of the Fe~6173~\AA~ line (at least for the electron beams with the average deposited energy flux $F_{d}\leq$\,5$\times$10$^{10}$\,erg\,cm$^{-2}$\,s$^{-1}$). However, significant changes in this line including a transient emission in the line core were found in the RADYN model calculated for an electron beam with a high (200 keV) low-energy cutoff, mimicking a deep penetration of the electron beam  \citep{Hong2018}.
	
	Because of the difficulties of explaining sunquakes, several alternative mechanisms were suggested. In particular, \citet{Hudson2008} and \citet{Fisher2012} suggested that the Lorentz force produced by a restructuring of the magnetic field during the flare energy release can provide a momentum sufficient for the excitation of sunquakes. \citet{Moradi2007} suggested that the low photosphere is heated by intense Balmer and Paschen continuum-edge radiation from the overlying chromosphere in white-light flares (the “back-warming” hypothesis). \citet{Sharykin2015a} noticed that some sunquake sources did not spatially coincide with the maxima of the hard X-ray emission and suggested that sunquakes could be produced by rapid dissipation of electric currents.
	
	The initiation of sunquakes by proton beams has been discussed in several papers \citep{Kosovichev2006,Moradi2007,Zharkova2007,Zharkova2015}. However, the gamma-ray emission produced by high-energy proton beams is rarely observed. Therefore, the observational results are inconclusive. \citet{Kosovichev2006} noticed that in the powerful flares of October 2003, the sunquake sources were close to both 50-100 keV hard X-ray and 2.2 MeV gamma-ray sources observed by the RHESSI satellite, but in the flare of July, 23, 2002, the sunquake sources were far away from the gamma-ray source but close to the hard X-ray sources \citep{Kosovichev2007}. Unlike the energetic electrons, the protons do not produce the hard X-ray radiation that would be consistent with observations \citep{Emslie1998}. Therefore, the evidence for the MeV protons has to rely on indirect diagnostics, such as impact linear polarization of chromospheric lines \citep{Henoux1990} and extreme Doppler shift of neutral hydrogen atom lines formed via charge exchange with the protons \citep{Orrall1976}. This effect, however, has not been observed yet \citep{Kerr2023}.
	
	Recently, \citet{Sadykov2023} developed flare radiative hydrodynamics (RADYN) simulations driven by proton beams that penetrate deeper than electron beams into the low atmospheric layers. They calculated synthetic Fe~I~6173~\AA~ line Stokes profiles and line-of-sight (LOS) observables of the SDO/HMI instrument as well as the 3D helioseismic response using a 3D acoustic model \citep{Stefan2020}. The results show the enhancement of the HMI continuum intensity and explain the generation of sunquakes.
	
	Here, we use the SDO/HMI spectro-polarimetric data to analyze the sunquake sources observed during the X1.5 flare of May 10, 2022, and compare the results with predictions of the flare hydrodynamics models. Section 2 describes the sunquake event.  Section 3 presents the results of the spectro-polarimetric analysis. In Section 4, we discuss the comparison with flare models. Our conclusions are presented in Section 5.
	
		\begin{figure} \centering \includegraphics[width=0.3\linewidth]{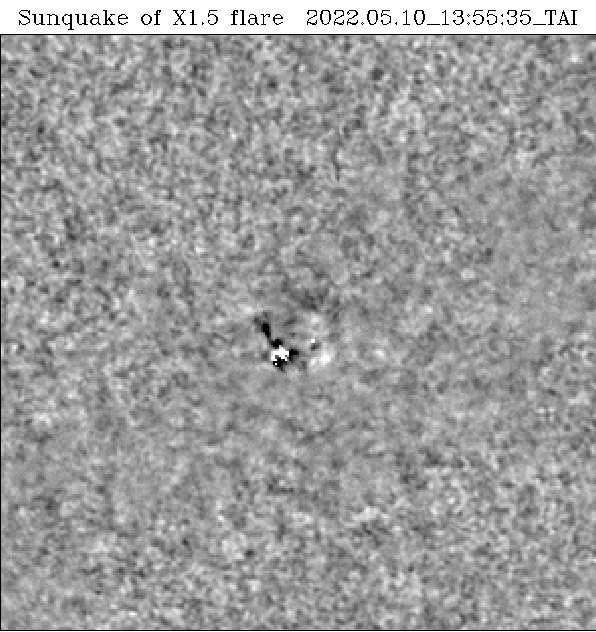} \caption{\label{fig1} Image of the online animation of the time series of the Dopplergrams observed by the SDO/HMI instrument in a 2mHz frequency band centered at 5 mHz. {The animated time series proceeds from 13:51\,UT to 14:48 on 2022/05/10. The animation reveals two sets of expanding wave ripples and indicates the presence of two sunquake sources.}} \end{figure}
	
	\section{Double Sunquake Caused by X1.5 flare of May 10, 2022}
	
	The first significant sunquake of Solar Cycle 25 observed during the X1.5 flare of May 10, 2022, revealed some interesting features that shed light on the origin of sunquakes. The animation constructed from a series of the 45-sec Dopplergrams filtered in a narrow frequency band around five mHz to isolate the acoustic waves from the low-frequency granulation noise shows initial impulsive perturbations and two sets of expanding ripples indicating the presence of two sunquake sources (Figure~\ref{fig1}).
	
	\begin{figure} \centering \includegraphics[width=0.6\linewidth]{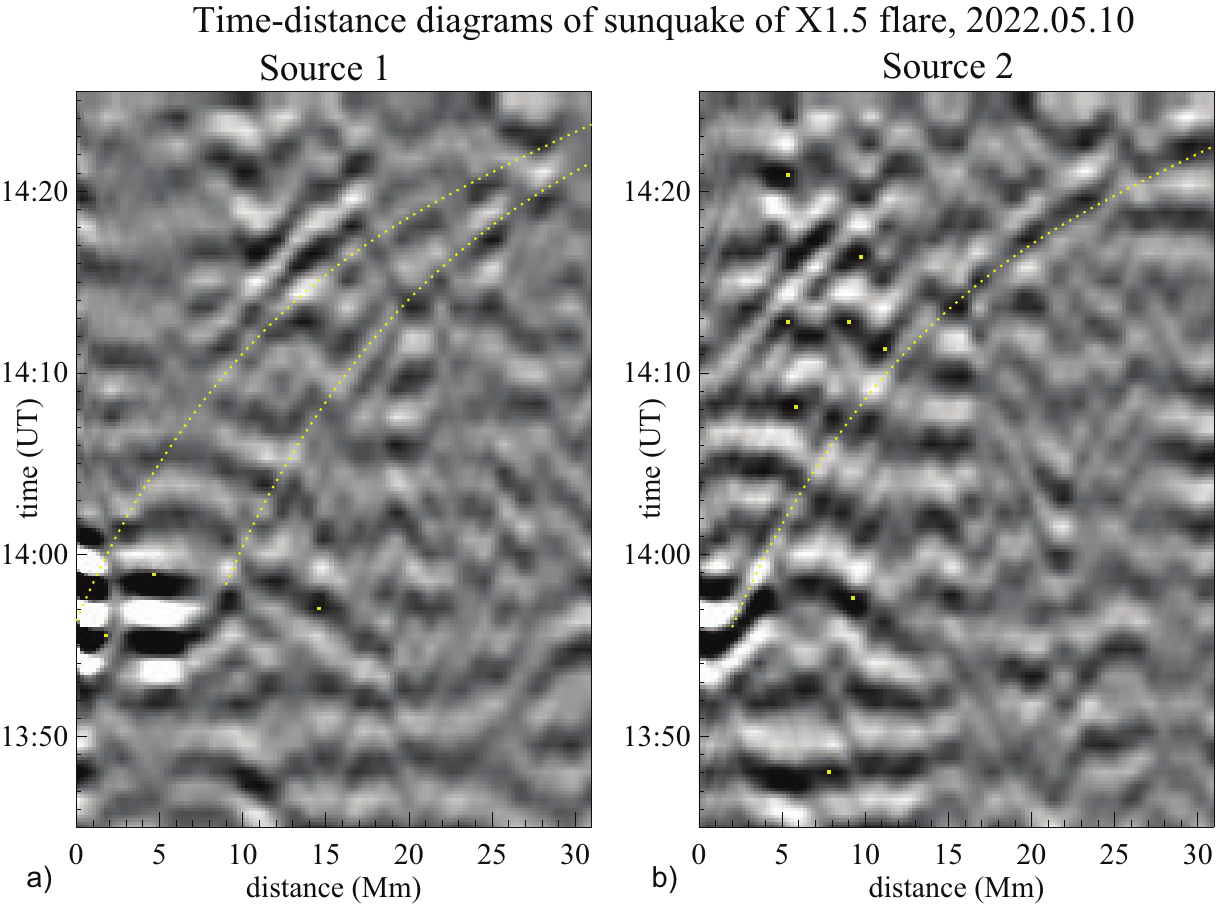} \caption{\label{fig2} The time-distance diagrams obtained by the angular averaging of the frequency-filtered Dopplergrams with the central points located in the Source 1 (a) and Source 2 (b) areas. The angular averaging is performed for the north-oriented quadrant.} \end{figure}

\begin{figure} \centering \includegraphics[width=0.5\linewidth]{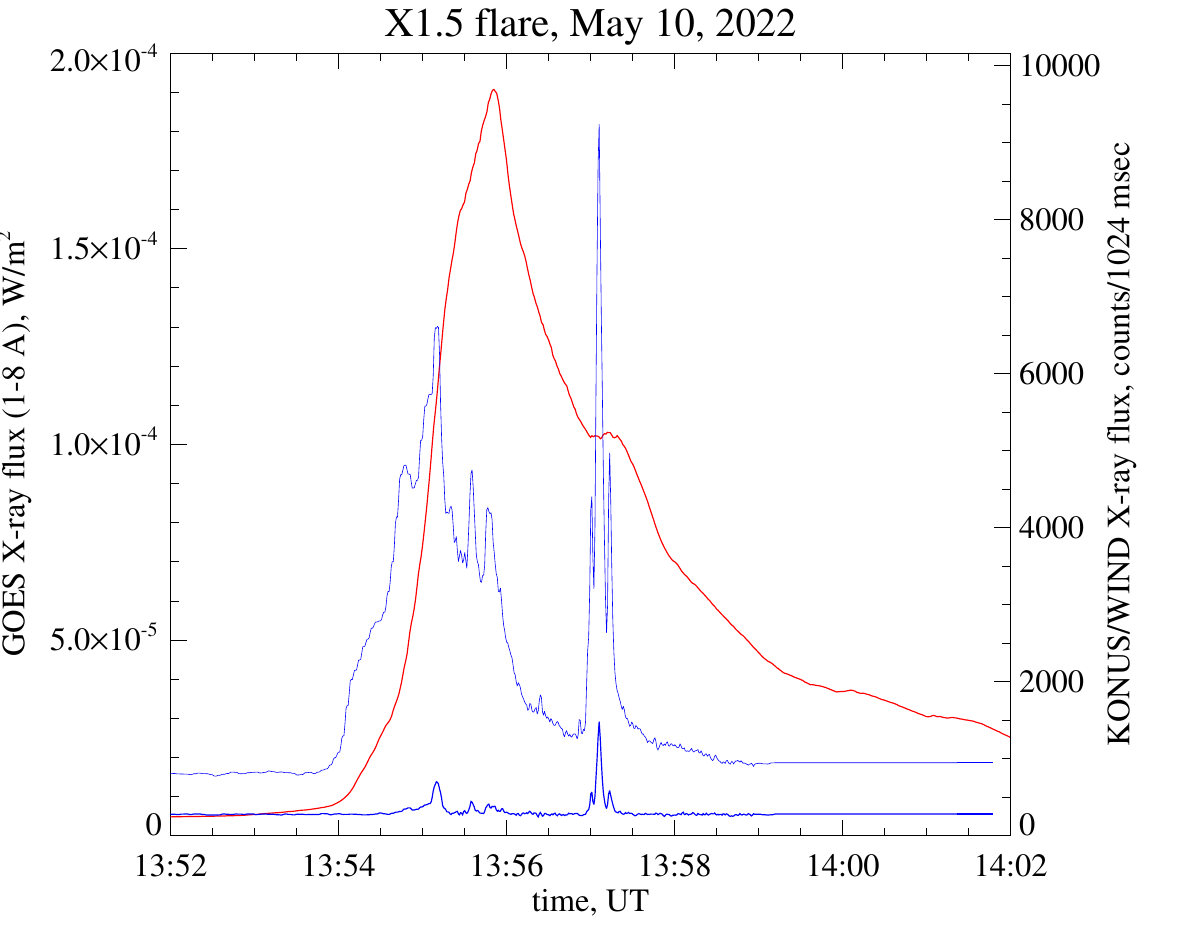} \caption{\label{fig3} The soft X-ray flare emission in the energy band 1.5-12 keV observed by the GOES satellite (red), and the hard X-ray emission in the energy bands 19-81 keV and 82-322 keV (blue) observed by the KONUS/WIND instrument \citep{Lysenko2022}.} \end{figure} 

\begin{figure} \centering \includegraphics[width=0.7\linewidth]{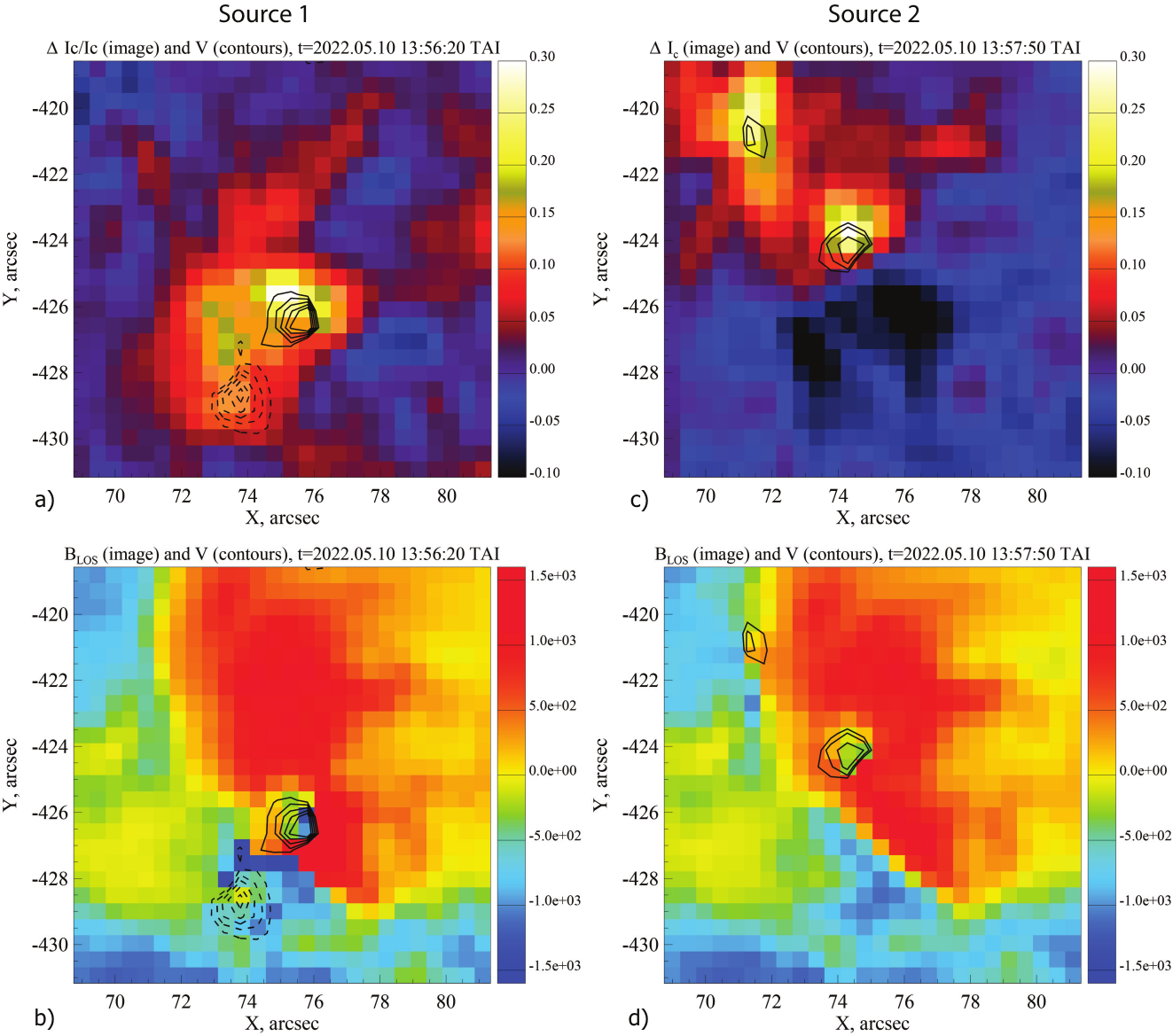} \caption{\label{fig4} The photospheric flare impacts observed by the HMI instrument around 13:56:20 UT (Source 1) and 13:57:50 UT (Source 2). Panels a) and c) show the relative continuum intensity enhancements during 45 sec of the impacts; panels b) and d) show the line-of-sight magnetograms during the flare impacts. The contour lines show the corresponding Doppler velocity signals for 3, 4, 5, 6, and 7 km/s. Solid lines show the positive (red-shifted) velocity, and dashed lines show the negative velocity.} \end{figure}

\begin{figure} \begin{center} \centering	\includegraphics[width=0.78\linewidth]{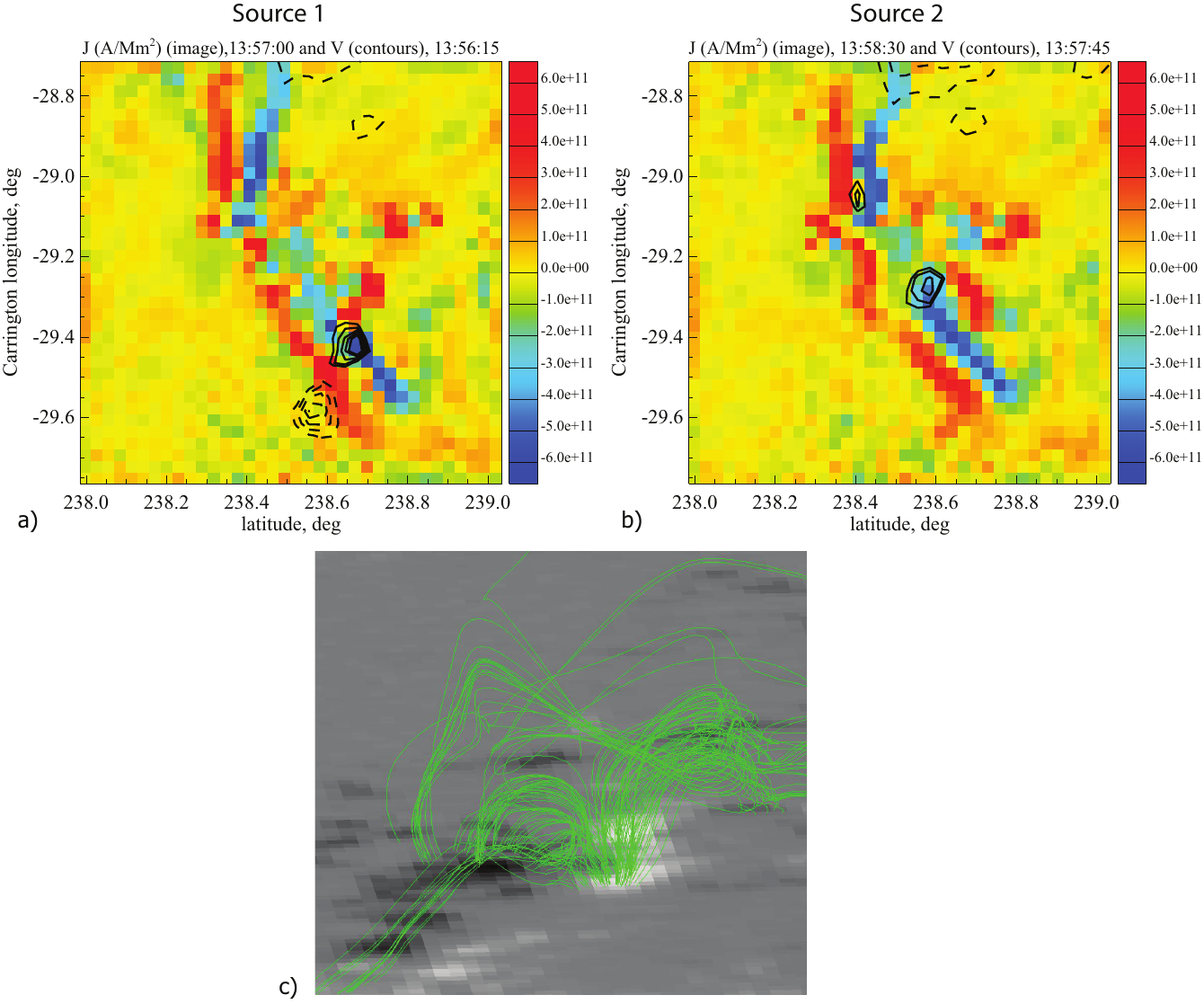} \caption{\label{fig5} The vertical electric currents at around a) 13:57:00\,UT (Source 1) and b) 13:58:30\,UT (Source 2). The overlaid contours of the Doppler velocity signals are the same as in Figure~\ref{fig4}. c) The magnetic field lines reconstructed in the NLFFF approximation from the HMI 720s vector magnetograms centered at 13:54 UT using the GX Simulator \citep[][the image is courtesy of G.Nita]{Nita2023}.} \end{center} \end{figure}

	The two double impact sources explain the double structure of the sunquake ripples appearing in the movie. The time-distance diagrams (Figure~\ref{fig2}) obtained by angular averaging of the ripples traveling in the North direction with the centers located in Sources 1 and 2 show a rather unusual double-ridge structure.
	
	Other interesting features of this flare are the presence of strong shearing flows in the polarity inversion line and a rapid restructuring of the magnetic field after the flare, resulting in the formation of a sunspot. In this respect, this flare resembles the M9.3-class flare of July 30, 2011, which, in addition to the helioseismic waves, caused a large expanding area of white-light emission and was accompanied by the rapid formation of a sunspot structure in the flare region \citep{Sharykin2015a}. Similarly, the flare produced hard X-ray emission in less than 300 keV energy band and no significant eruption.

	Indeed, the hard X-ray flux observed by the Konus/Wind instrument \citep{Lysenko2022} shows two emission peaks at 13:55 UT and 13:57 UT. Interestingly, there was a 1-min gradual phase in the 19-81 keV X-ray emission before the main first peak (Figure~\ref{fig3}).
	
	Because the flare was located near the central meridian, the photospheric perturbations associated with the hard X-ray impulses are well-resolved in the HMI images taken around 13:56 UT and 13:58 UT \citep[each HMI observing sequence takes 45 sec,][]{Couvidat2012a}. The top panels of Figure~\ref{fig4} show the relative difference between two consecutive continuum intensity images taken at 13:55:35 UT and 13:56:20 UT (Source 1; left panel) and 13:57:05 UT and 13:57:50 UT (Source 2; right panel). The contour lines show {the line-of-sight velocity perturbations calculated from the HMI 45-sec cadence Doppler-shift data} at 13:56:20 UT and 13:57:50 UT for 3, 4, 5, 6, and 7 km/s (solid curves show positive red-shifted velocity; negative blue-shifted velocity is shown by dashed). The bottom panels show the line-of-sight magnetograms at the exact times as the Doppler velocity signals.

	These images show that each of the sunquake sources had a double impact structure, most likely corresponding to the footpoints of magnetic flux tubes across the polarity inversion line. The distance between the impacts of Source 1 is about 2-3 arcsecs, indicating the energy release in a low-lying magnetic flux tube. The distance between the impacts of Source 2 is about 4 arcsecs; however, it appeared stretched along the polarity inversion line. The impacts are located in regions of strong vertical electric currents calculated from the HMI vector magnetograms (Figures~\ref{fig5}(a) and (b)). These images show that the electric currents flow across the polarity inversion line along compact magnetic loops, forming an arcade. The sequence of HMI images shows that the flare impacts evolve in a wave-like fashion along the arcade, starting from the Source 1 location and ending at Source 2. The complex magnetic field structure with low-lying magnetic field lines connecting the opposite polarities is evident from the non-linear force-free reconstruction (Figure~\ref{fig5}(c)) using the GX Simulator \citep{Nita2023}.
	
	It should be noted that the HMI observing scheme may not provide accurate velocity and magnetic field measurements during the impacts because the line profile may be significantly distorted \citep[e.g.][]{vSvanda2018}. Nevertheless, these perturbations are substantially stronger than the electron-beam RADYN models in the F-CHROMA database predict \citep{Sadykov2020}. Strong photospheric perturbations are typical for sunquake events.
	
	\section{Spectro-Polarimetric Characteristics of Sunquake Sources}
	
	The HMI instrument performs spectro-polarimetric measurements by taking linear and circular polarized narrow band images in six positions across the Fe~I~6173~\AA~ line with two cameras \citep{Couvidat2012a,Schou2012}. Camera 1 observes the linear polarized light, and Camera 2 measures the circular polarized light. The 4096x4906 pixel images cover the whole disk with the 0.5 arcsec/pixel sampling. The spatial resolution of the instrument is 1 arcsec.  The polarized images are taken every 3.75 sec, and the whole spectro-polarimetric measurement sequence takes 90 seconds for the linear polarization and 45 sec for the circular polarization. The polarized images are sequenced in an order minimizing systematic errors caused by the line variations during the observing cycles and combined into the Stokes line profiles.
	Thus, the HMI observations provide the full Stokes line profiles for the Fe~I~6173~\AA~ line with 78~m\AA~ spectral, 1 arcsec spatial, and 90-sec temporal resolution. In the standard HMI data processing pipeline, the Stokes profiles are used to determine the vector magnetic field and line-of-sight velocity by applying a simplified Milne-Eddington inversion procedure. However, it does not provide robust results for flare properties because of the rapid variations of the line profile. Therefore, we focus on analyzing the Stokes profile variations in the sunquake sources. 

	To investigate the temporal behavior of the sunquake sources, we remapped the Stokes profile images onto the heliographic coordinates and tracked the local patches of these images with the local differential rotation rate. Then, the regions of the flare impacts are identified as compact short-duration high-intensity brightenings in the Stokes~I images. The characteristic size of these impacts does not exceed 0.1-0.2 heliographic degrees (1-2~Mm), which is at the limit of the HMI spatial resolution. The duration of these impacts does not exceed 2-4 min. The time profiles are characterized by fast growth and slower decay. After the impacts, some of the Stokes profiles returned to their pre-impact properties, but in all cases, we see irreversible changes either in the circular or linear polarization profiles.

\begin{figure} \centering \includegraphics[width=0.8\linewidth]{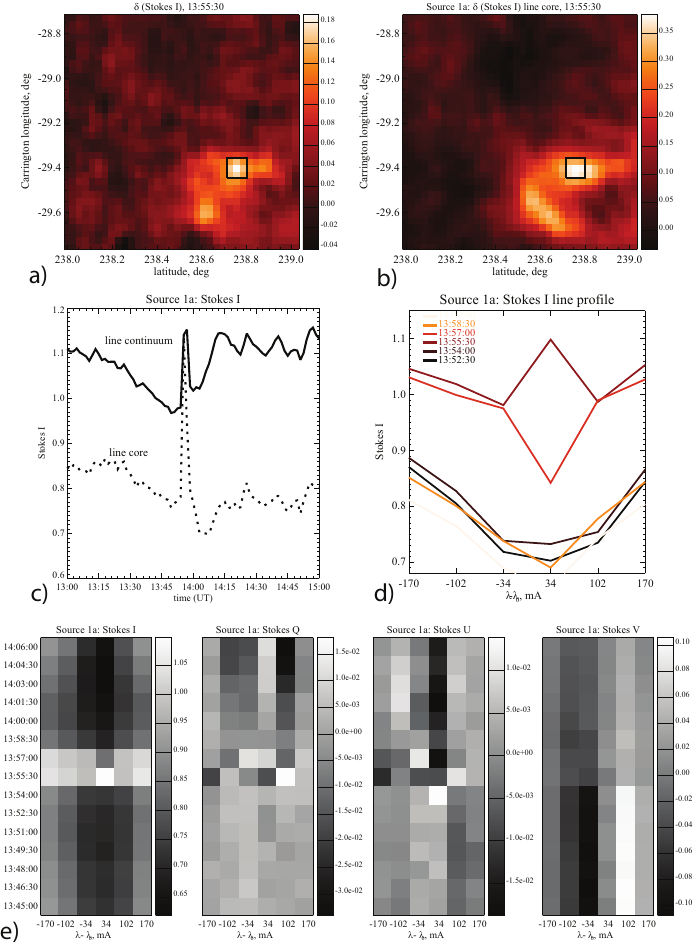} \caption{\label{fig6} Source 1a: a) image of the Fe~I~6173~\AA~ line continuum intensity time difference during the flare impact; b) image of the line core intensity time difference during the flare impact; c) the temporal behavior of the line continuum (solid curve) and core (dashed curve) averaged over the black box in panels a) and b); d) variations of the line profile at different time moments before, during and after the impact; e) the corresponding variations of Stokes I, Q, U, and V line profiles. } \end{figure}

\begin{figure} \centering \includegraphics[width=0.85\linewidth]{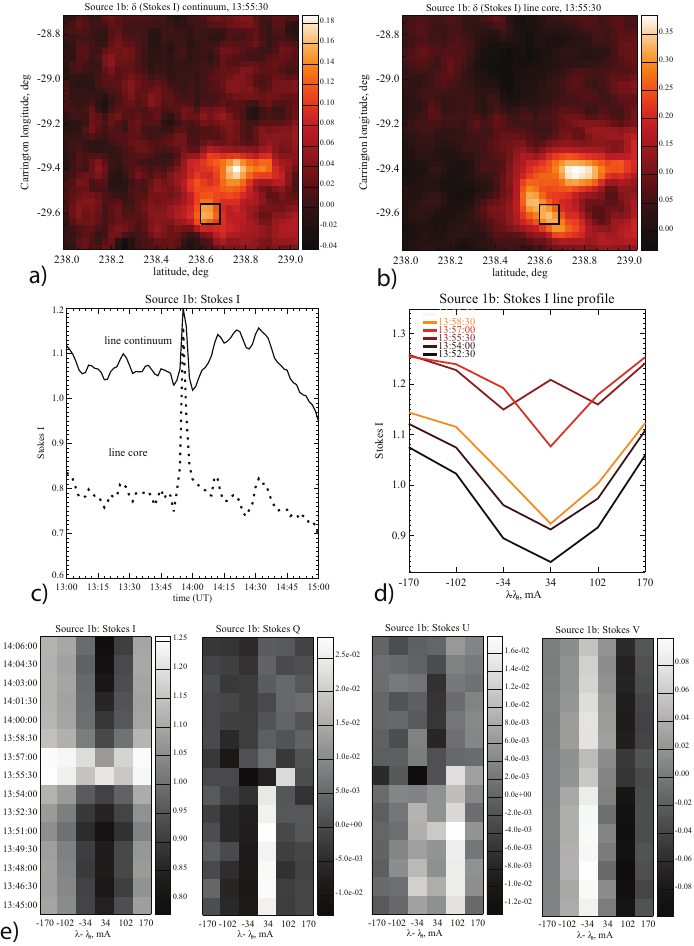} \caption{\label{fig7} The same as in Figure~\ref{fig6} for Source 1b. } \end{figure}

\begin{figure} \centering \includegraphics[width=0.85\linewidth]{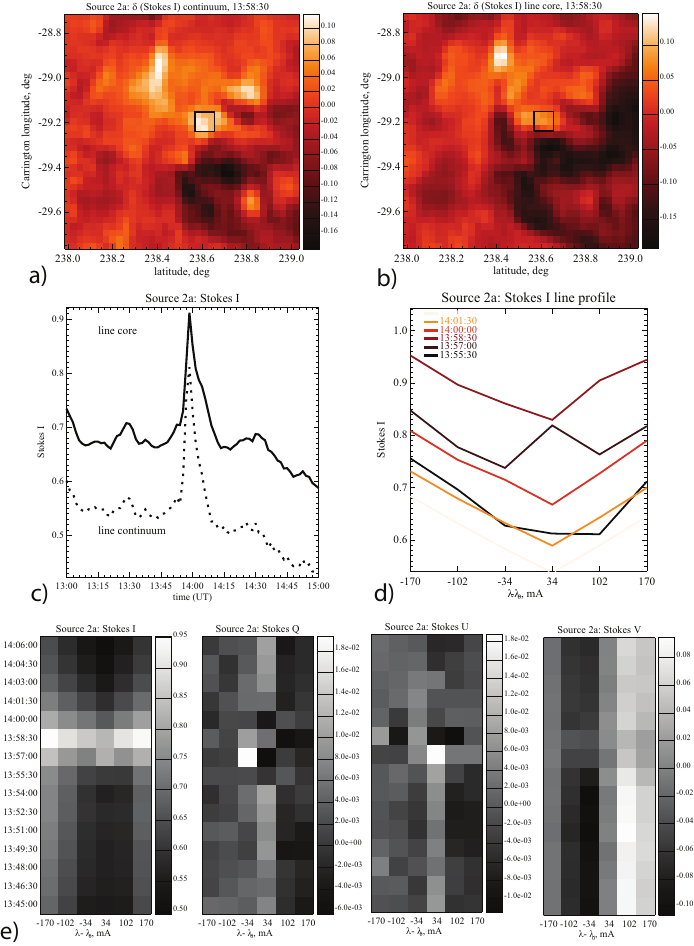} \caption{\label{fig8} The same as in Figure~\ref{fig6} for Source 2a. } \end{figure}

\begin{figure} \centering \includegraphics[width=0.85\linewidth]{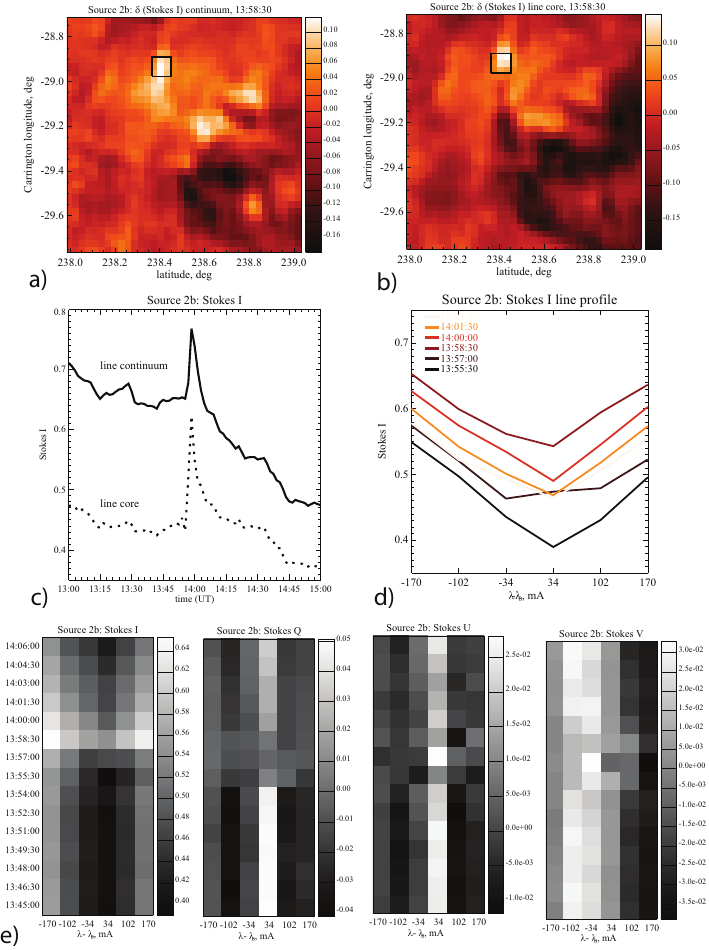} \caption{\label{fig9} The same as in Figure~\ref{fig6} for Source 2b. } \end{figure}

	Figures~\ref{fig6}-\ref{fig9} show the filtergram images of the sunquake sources in the line continuum and the line core (panels a-b), the temporal variations in the line continuum and core (panels c), averaged over $3\times 3$-pixel areas marked by black squares, the Stokes~I line profiles at various moments of time (panels d), and the time-wavelength diagrams of the Stokes profiles (panels e). All Stokes profiles show substantial rapid variations during two or three 90-sec HMI Camera 1 observing sequences. The line intensity increases by about 10-30\% in the line continuum (far line wings) and by 30-50\% in the line core. In the line core, the peak intensity is reached earlier than in the line continuum, indicating that the impact affected higher layers of the solar photosphere earlier than the lower layers. However, the exact time lag cannot be determined because of the 90-sec Stokes profiles measurement cycle. At the peak values, the line core was in emission in three cases (Sources 1a, 1b, and 2a). Before the initial impacts (Sources 1a and 1b), the Stokes I showed a brief ~3-min darkening in relatively large areas, which are also noticeable in the Stokes~I images in Figures~\ref{fig6}(d)-\ref{fig9}(d). The temporal darkening may be related to the heating of photospheric layers in the flare pre-impulsive phase, which results in an increase of the negative hydrogen opacity (due to increased ionization) and a temporary decrease of the continuum flux \citep{Grinin1983,Henoux1990a}.
	
	After the impulsive variations, the Stokes~I profiles quickly return to their pre-flare values, indicating the rapid cooling of the photospheric plasma. However, the Stokes Q, U, and V profiles, characterizing linear and circular polarization, show permanent or long-duration changes (Figures~\ref{fig6}(e)-\ref{fig9}(e)), indicating changes in the magnetic field geometry and strength. Specifically, in Source 1a, the values of Stokes Q and U increased while the Stokes V decreased, indicating an increase in the horizontal field strength and a decrease in the vertical field strength. In Source 1b, the initial strong Stokes Q and U permanently decreased, while Stokes V returned to the pre-flare values. In Sources 2a and 2b, the Stokes V profiles returned to the pre-flare values, and Stokes Q and U experienced relatively weak permanent changes. The permanent changes could be due to the height shift of the line formation region because of the flare heating. However, the Stokes I profiles in all cases quickly returned to the pre-flare values, meaning that the heating did not last long. Thus, the flare impacts are probably associated with fast permanent changes in the magnetic field strength and topology.
	
\begin{figure} \centering \includegraphics[width=0.85\linewidth]{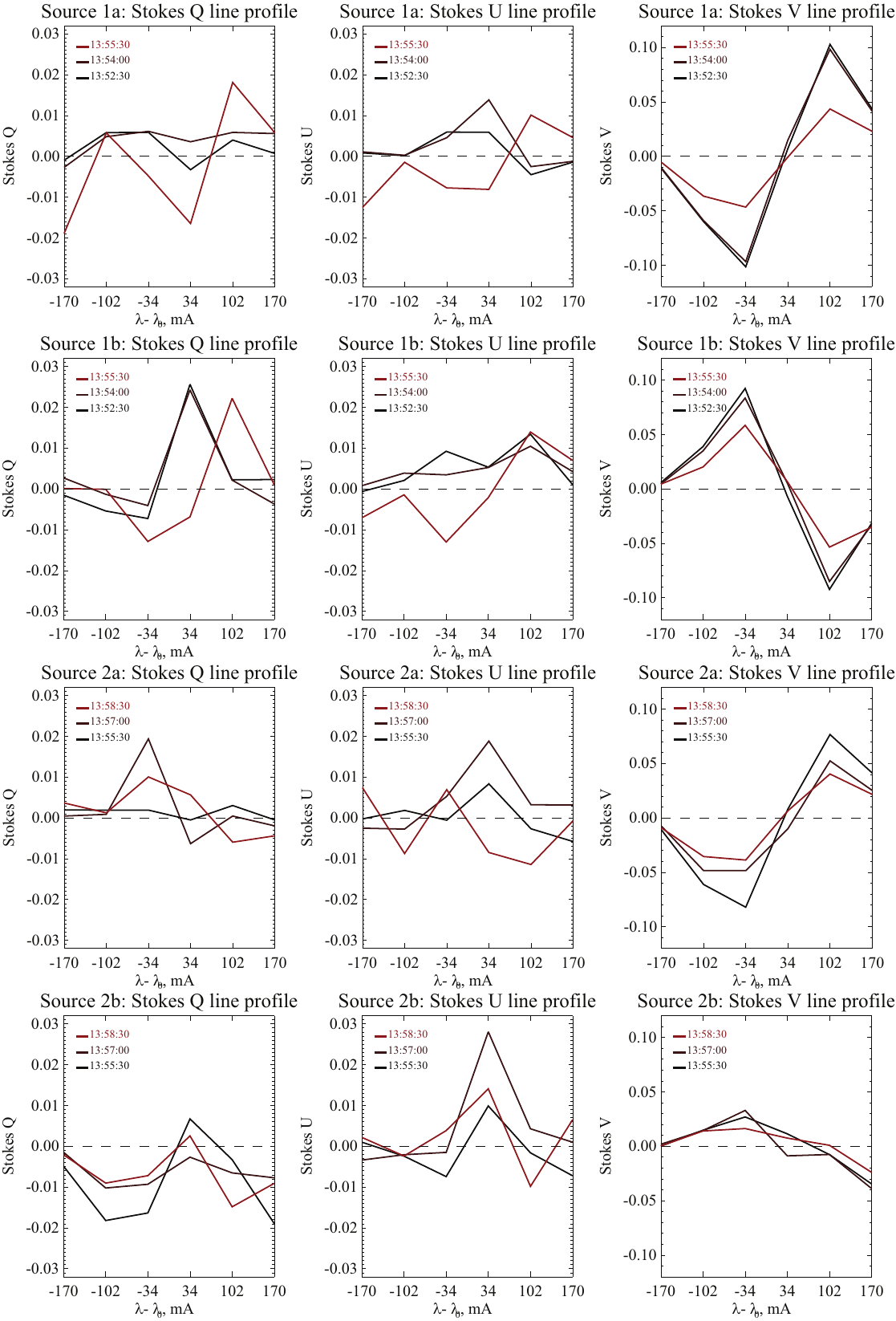} \caption{\label{fig10} The evolution of the Stokes Q, U, and V profiles in the flare sources before and during the Stokes I emission peaks. } \end{figure}

	The observed transient changes in Stokes Q, U, and V are likely affected by fast non-equilibrium variations of the plasma properties. {Figure~\ref{fig10} shows the variations of Stokes Q, U, and V profiles for three consecutive measurements taken just before the Stokes I peaks, shown in Figures~\ref{fig6}(c)-\ref{fig9}(c), and during the peaks. The observing times are indicated in Figure~\ref{fig10} panels. The Stokes I magnitude slightly decreased in all sources, but Stokes Q and U, representing the linearly polarized components, vary significantly. The origin of these transient variations is unclear but may be related to impact linear polarization, previously discussed for chromospheric lines \citep{Henoux1990,Simnett1995}.}
	
	\section{Physical Origin of Sunquake Sources}
	
	The standard flare model assumes that the solar atmosphere is heated by beams of electrons accelerated to high energies in the corona. However, previous studies have shown that the electron beams and their hydrodynamic effects are possibly too weak to explain the HMI observations of the photospheric impacts producing dramatic changes in the Fe~I 6173~\AA~ line and the helioseismic waves \citep{Sadykov2020}. Only the electron-beam models with the high low-energy cut-off of 200~keV produced a line-core emission \citep{Hong2018}.
	
\begin{figure} \centering \includegraphics[width=0.7\linewidth]{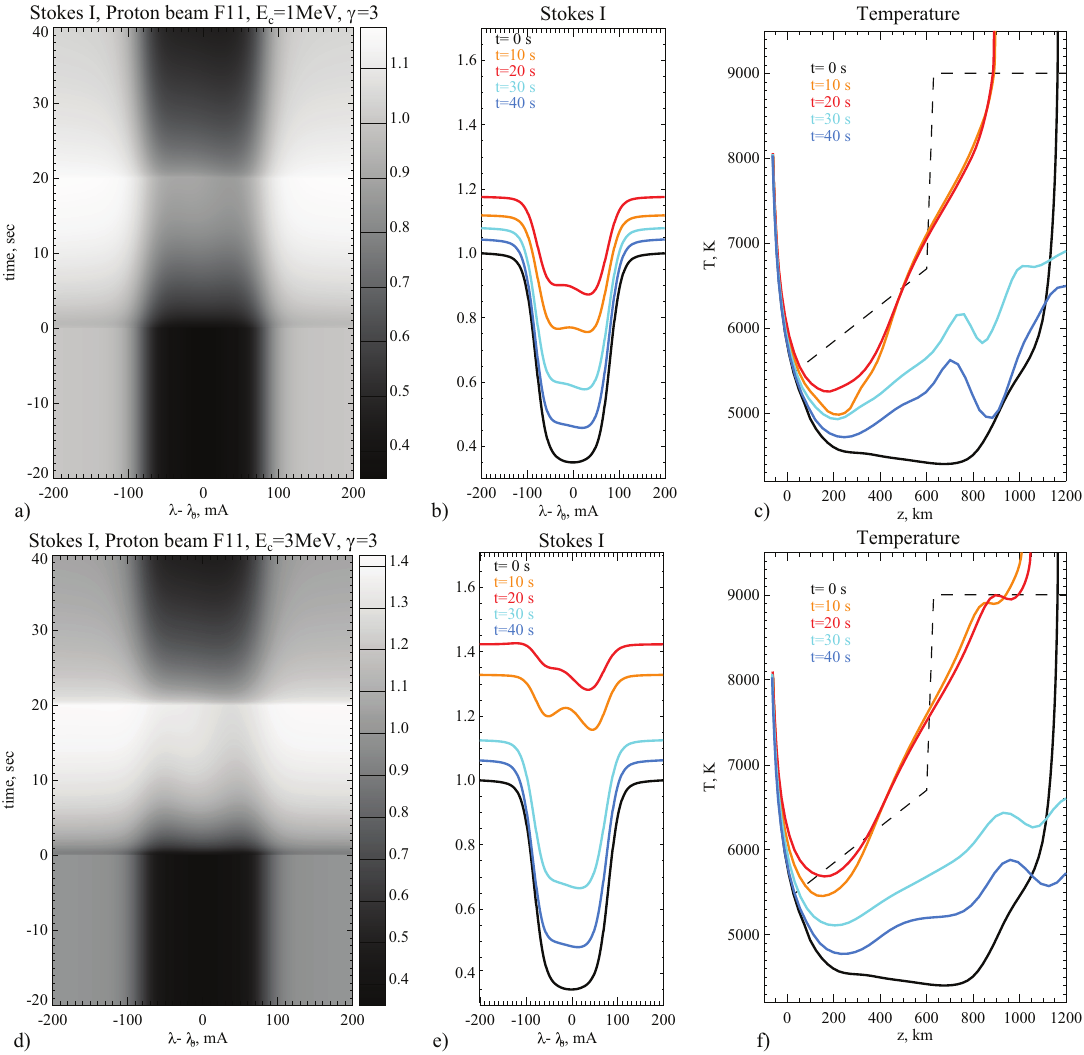} \caption{\label{fig11} (a) Variation of the Stokes I, (b) line profiles at various time moments before, during and after the impact, and (c) the corresponding temperature profiles in the low atmosphere, calculated in the radiative hydrodynamics (RADYN) model driven by a proton with the energy flux of $10^{11}$ erg$\,$cm$^{-2}\,$s$^{-1}$, the power law index $\gamma=3$, and the low energy cut-off of $E_c=1$~MeV; (d-f) the same as panels (a-c) for $E_c=3$~MeV \citep{Sadykov2023}. The dashed line shows the semi-empirical model of a white-light flare, obtained by \citet{Kleint2016} } \end{figure}

	Recent modeling of proton beams in the framework of the flare hydrodynamics RADYN model showed that the proton beams with a typical energy flux of $10^{11}$ erg$\,$cm$^{-2}\,$s$^{-1}$ (the average energy flux just twice higher than in FCHROMA), the power law index $\gamma=3$, and relatively high low cut-off energies $E_c=1-3$~MeV can heat the photospheric layers to high temperatures sufficient to explain the observed continuum intensity increase and the line-core emission \citep{Sadykov2023}. Figure~\ref{fig11} shows the temporal variations of the Stokes I profiles calculated for two RADYN models for the proton beams with the low cut-off energies of $E_c = 1$~MeV (panels a and b) and 3~MeV (panels d-e), as well as the temperature profiles in the low chromosphere and photosphere. The duration of the beam heating was 20~s. Both cases show variations that are qualitatively similar to the observed variations, including the line core emission, which is more decisive for the 3~MeV case. {Although, the observed line core emission is stronger than in the models. In addition, the line profile becomes shallow. The line core depression (the line depth measured at the Fe\,I\,6173\,{\AA} rest wavelength relative to the continuum near the line), averaged over the 20-sec beam heating period, is about 0.34 for the 1~MeV case and 0.13 for 3~MeV. For comparison, the observed line core depression averaged over the time intervals corresponding to the flare emission peaks (Figures~\ref{fig6}(c)-\ref{fig9}(c)), is about 0.11 in Sources 1a and 2a, 0.14 in Source 1b, and 0.20 in Source 2b.}   

\begin{figure} \centering \includegraphics[width=0.6\linewidth]{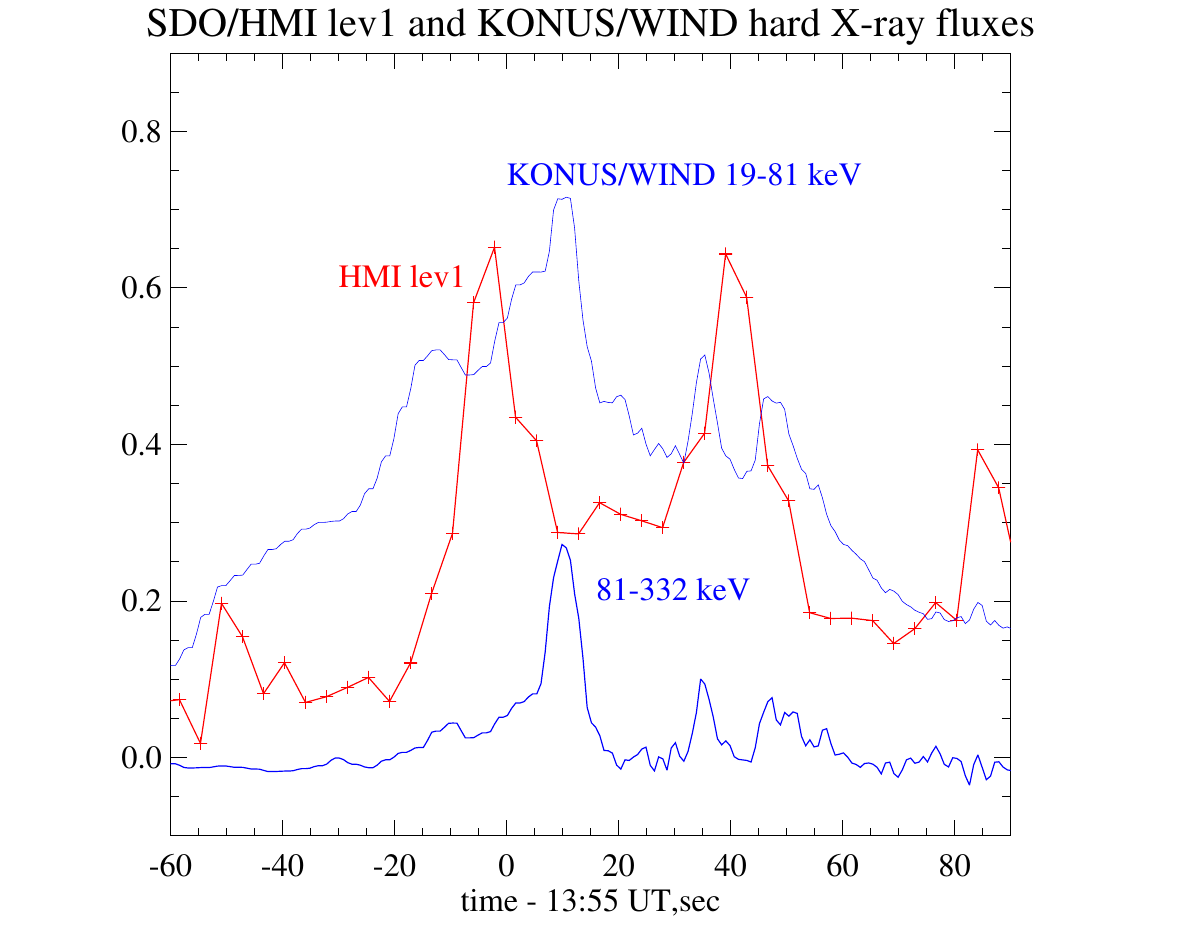} \caption{\label{fig12} Red curve shows the filtergram signal at Source 1a from the HMI Camera 2 (level 1 data) normalized to the corresponding pre-flare values observed at 13:30 UT. The blue curves show the hard X-ray signal observed by the KONUS/WIND instrument in the energy range of 19-81 keV (thin curve) and 81-332 keV (thick curve). } \end{figure}
	
	The line core emission is apparently due to the stronger heating in the temperature minimum regions between the photosphere and chromosphere. However, such beams do not generate helioseismic waves of the amplitude observed in sunquakes. \citet{Sadykov2023} found that the proton beams with $E_c \leq 100$~keV can explain sunquakes because of the higher momentum impact, but the heating by such beams is too weak to explain the Fe~I 6173~\AA~ line variations. The dashed lines in Figures~\ref{fig11}(c) and (f) show the semi-empirical model derived by \citet{Kleint2016} for a white-light emission kernel of the X1 flare of 2014, March 29. The semi-empirical model is fairly close to the 3~MeV proton beam model. This confirms their conclusion that both the chromosphere and photosphere are strongly heated in white-light flares.
	
	Protons are likely accelerated at the same time as electrons. Unlike the protons, the accelerated electrons produced hard X-ray emission, which was observed by the KONUS/Wind instrument. In this respect, it is important to consider the exact timing of the hard X-ray and photospheric impulses. To investigate this relationship, we used the filtergrams obtained in alternating right and left circular polarizations from the HMI Camera 2 with a 3.75~s cadence. To reduce the variations caused by the spectral line scanning, we normalize these data to the corresponding Camera-2 data taken 20 min before the impacts. In Figure \ref{fig12}, the normalized HMI signal for Source 1a (red curve) is compared with the hard X-ray emission from the KONUS/WIND instrument. The HMI signal starts simultaneously with the hard X-ray emission but reaches the maximum about 15 seconds earlier than the X-ray emission, particularly in the 81-332 keV energy range.
	
	This fact, as well as the absence of measurable X-ray emission with energies greater than 300 keV, do not support the suggestion \citep{Wu2023} that sunquakes could be initiated by high-energy electrons. Previously, sunquake sources were associated with CME eruptions, impulsive Lorentz force, and electric currents. In this flare, there was no eruption. An impulsive Lorentz force cannot provide such strong and fast heating in the low photosphere where the plasma beta parameter is greater than unity. The dissipation of electric current would require their filamentation, e.g., due to the Joule-overheating instability \citep{Sokolov1978}, which may not be developed on such a short time scale unless the electrical resistivity is enhanced due to turbulence or other effects \citep{Heyvaerts1974}. 
	
	Therefore, we conclude that the momentum impact caused by deeply penetrating low-energy proton beams is the most viable mechanism of sunquake generation. However, such beams do not explain the observed emission in the Fe~I 6173~\AA~ line core because of the weak heating of the low atmosphere. The high-energy electron and proton beam models can qualitatively explain the line-core emission, but the models do not explain sunquakes.
	
	\section{Discussion and Conclusion}
	
	Significant variations of the photospheric Fe~I 6173~\AA~ line, observed by the HMI instrument onboard SDO during white flares and sunquake events, were noticed before \citep{MartinezOliveros2011}. The sunquake event observed during the X1.5 flare of May 10, 2022, developed in a relatively simple magnetic region located close to the central meridian, provided an opportunity to perform a detailed spectro-polarimetric analysis and compare with the predictions of the radiative hydrodynamic (RADYN) models. The HMI filtergrams obtained in six positions across the spectral line provided good coverage of the line profile because of the flare location in the central part of the solar disk.
	
	The photospheric flare impacts are identified as sharp variations of the HMI observables: continuum intensity, Doppler velocity, and line-of-sight magnetic field. The impacts appeared in pairs across the magnetic polarity inversion line, separated by 2-3 arcsec, indicating that they represent footpoints of low-lying magnetic loops. The impacts started in the southern part of the active region simultaneously with the hard X-ray impulses and propagated along the polarity inversion line with a speed of about 50-60 km/s. The most significant impacts were separated by about 6-8 arcsec with a time lag of about 2 min, and these impacts resulted in the excitation of two partially overlapping helioseismic waves. Both waves are well-resolved in the time-distance diagrams. This is the first case of a `double sunquake' known to us.
	
	We analyzed the temporal evolution of the four strongest impacts, using the line Stokes profiles reconstructed from the HMI linear and circular polarized filtergrams with 90-sec cadence. The results showed fast variations with rapid growth and slower decay lasting 3-4 min, coinciding with the hard X-ray impulses. The variations in the line core appeared slightly ahead of the variations in the line wings, indicating that the heating started in the higher atmospheric layers and propagated downward. The sharp brightenings were preceded by darkening lasting 3-5 min, particularly noticeable in the initial impact area (Sources 1a and 1b).
	
	In all impacts, the Stokes I profiles returned to their initial values, meaning a rapid cooling of the heated plasma. However, some Stokes Q, U, and V profiles showed permanent changes, indicating irreversible changes in the magnetic field strength and geometry. The HMI magnetograms showed substantial rapid restructuring of the active region magnetic field, resulting in the formation of a sunspot. A similar process occurred after a sunquake flare of July 30, 2011 \citep{Sharykin2015a}.
	
	The most prominent feature of the line profile variations is the emission in the line core in three of the four sources. It indicates strong impulsive heating in the low chromosphere and photosphere. The radiative hydrodynamics (RADYN) simulations of \citet{Sadykov2023} showed that the proton beams with an energy flux of $10^{11}\,$erg$\,$cm$^{-2}\,$s$^{-1}$, the power law index $\gamma=3$, and the low energy cut-off $E_c=1-3\,$MeV can explain the emission in the Fe~I 6173~\AA~ line core. Still, the sunquake amplitude, observed by the HMI instrument, was obtained only for the models with $E_c \leq 100\,$keV.  Using the HMI level 1 filtergrams, obtained with a 3.75~s cadence, we determined that the initial photospheric impact started simultaneously with the hard X-ray impulses but reached the maximum 10-15 sec earlier than the hard X-ray fluxes.
	
	Putting all these results together, we come to the suggestion that the flare energy release, which occurred in low-lying magnetic loops, produced high-energy electrons and protons. The electrons, precipitating into the high chromosphere, produced hard X-ray emission at the loop footpoints, but the protons penetrated deeper into the photospheric layers and delivered the momentum sufficient for the generation of sunquakes. The Fe~I 6173~\AA~ line core emission indicates strong impulsive heating of the lower chromosphere and photosphere, to which high-energy protons ($E \gtrsim 5$ MeV) and electrons ($E \gtrsim 200$ keV) could contribute. However, a quantitative model consistently explaining these phenomena and the observed variations of the line polarization is still lacking.

	\section*{\textsc{Acknowledgments}} \noindent  \normalsize This research was supported by NSF grant 1835958 and NASA grant NNX14AB68G. VMS acknowledges the NSF FDSS grant 1936361. We thank Joel Allred and Adam Kowalski for providing us with the RADYN proton and electron beam heating models used in this study.

\bibliographystyle{aasjournal}

\end{document}